\pdfoutput=1
\documentclass[conference,final,a4paper]{IEEEtran}
\usepackage[utf8]{inputenc}
\usepackage{enumerate}
\usepackage{xcolor}
\usepackage{algorithm,algorithmic}
\usepackage{amsmath}
\usepackage{amssymb}
\usepackage{mathtools}
\usepackage{comment}
\usepackage{multicol}
\usepackage{lipsum}
\usepackage[caption=false, font=footnotesize]{subfig}

            \author{Amartya Sanyal\IEEEauthorrefmark{3},\IEEEauthorblockN{Sanjana Garg\IEEEauthorrefmark{1},Asim Unmesh\IEEEauthorrefmark{2}}
             \IEEEauthorblockA{Dept. of Computer Science And Engineering\\
              Indian Institute of Technology, Kanpur\\
              Email:
              \IEEEauthorrefmark{3}amartya18x@gmail.com
              \IEEEauthorrefmark{1}garg.sanjana.95@gmail.com,
              \IEEEauthorrefmark{2}a.unmesh@gmail.com
            }
      (All the authors have equal contributions and the order is random)}
\title{Agent based simulation of the evolution of society as an alternate maximzation problem}
\begin{document}

\maketitle
\begin{abstract}
 Understanding the evolution of human society, as a complex adaptive system, is a task that has been looked upon from various angles. In this paper, we simulate an agent-based model with a high enough population tractably. To do this, we characterize an entity called \textit{society}, which helps us reduce the complexity of each step from $\mathcal{O}(n^2)$ to $\mathcal{O}(n)$. We propose a very realistic setting, where we design a joint alternate maximization step algorithm to maximize a certain \textit{fitness} function, which we believe simulates the way societies develop. Our key contributions include (i) proposing a novel protocol for simulating the evolution of a society with cheap, non-optimal joint alternate maximization steps (ii) providing a framework for carrying out experiments that adhere to this joint-optimization simulation framework (iii) carrying out experiments to show that it makes sense empirically (iv) providing an alternate justification for the use of \textit{society} in the simulations.
\end{abstract}

\section*{Introduction}
The purpose of this work is to develop a framework for modeling the evolutionary dynamics of a society populated with heterogeneous agents. Though history has often tried to portray society as a coherent organization trying to optimize a global objective, society has often yielded agents bound together by constraints(laws) trying to reach their local optimum. What is worthwhile to notice is that, just like language, societies that have developed independently in different parts of the world have developed similar characteristics in terms of organization. This might have been because societies that worked in a different way failed to survive the sands of time or because societies, somehow, ended up developing in a similar way. Either way, this suggests that \textit{successful} societies were bound with certain constraints just like the theory of \textit{Universal Grammar} and one of them, as have been proposed by Darwin\cite{darwin1951origin}, is the survival of the fittest.

In other words, all individuals try to maximize their fitness lest they should perish and all societies try to maximize their \textit{happiness} lest they should perish. We propose that these objectives are related to each other but not in a trivial sense. In a hugely complex interactive heterogeneous system such as a society, this cannot be tractably reduced in complexity than a complex adaptive system\cite{miller2009complex}. In this system, all agents are motivated to maximize their well being as much as possible. For this text, we will refer to this well being as \textit{happiness}. It is to be noted that the people are not given a free hand in this optimization step but are often bounded by constraints imposed by society and the natural environment. Constraints imposed by societies include the  presence of norms and those induced by environment include climates and geographical features. In this paper, we model these two optimization steps, we develop the characteristics of the agents of this CAS and we also model the society.

Interaction in a society intuitively refers to the $\mathcal{O}(n^2)$ pairs of interactions possible. An alternative way of looking at the characteristics of the society is to look at it as a facilitator for a  mathematical simplification of a more expensive computational problem. The \textit{happiness} of a society is an aggregation of the $\mathcal{O}(n^2)$ interactions possible in a society. We present the society as a latent variable modeling these $\mathcal{O}(n^2)$ in just $\mathcal{O}(n)$ interactions here the society is a constant participant of all these two-person interactions.

Though it has become common to assume rationality of people in the study of social sciences\cite{scott2000rational,downs1994conflict}, we align with the view that human decision making is constrained when it comes to rationality. This is not only due to the fact than an average human is not always able to figure out the most rational action due to limits in knowledge and cognitive ability but also because, as we have emphasized before, his objective and the society's objective may not be aligned. In other words, the agents follow the idea of \textit{bounded rationality}\cite{simon1957models} and our model tries to ensure this when it looks upon decision making as an optimization step. Instead of optimizing the \textit{happiness} function, we rather create an augmented augmented fitness function, which inherits this idea of \textit{bounded rationality}.
In Section 2, we introduce the problem setting, where define the terms and notations. In section 3, we look into the protocol for the simulation itself. In section 4, we look at our experimental simulations and the results and then in section 5, we finally conclude with our observations and our ideas regarding future work.

\section*{Problem setting}
In our model, we have two agents population and society.
\begin{itemize}
    \item $\mathcal{A}$ - \{Population,Society\}  
    \item $\mathcal{S}_p$ - Set of pure strategies of population which are essentially the characteristics of an individual that we have considered in our model.
    \item $\mathcal{S}_s$ - Set of pure strategies of society which represent characteristics of a society.
Let us denote each of the strategies as follows:
    \item $\mathcal{S}_p = \{a,b,c,d,e,f,g,h\}$
    \item $\mathcal{S}_s = \{1,2,3,4,5,6,7,,8,9,10,11,12,13\}$
\end{itemize}
We look at the following individual traits while modelling the agents$ \{a,b,c,d,e,f,g,h\}$ : \textit{Intellectual/Education level, Physical Strength, Obedience, Flexibility towards change,Health/Immunity, Sincerity towards work, Family-oriented, Religious}\\
The society is modellied with the following traits$\{1,2,3,4,5,6,7,,8,9,10,11,12,13\}$:
\textit{Literacy rate/Education levelL, iving standards/income, Crime rate, Agrarian, Industrial, Conservati, Communist}
\subsection*{Problem modeling}
\begin{itemize}
\item $\sigma_s$ - mixed strategy of society
\item $\pi(\sigma,\sigma_s)$ - payoff for individual adopting mixed strategy $\sigma$
\item $\mathcal{I}_{|\mathcal{S}_p|\times|\mathcal{S}_s|}$ - Interaction between society and population
\item $\pi(\sigma,\sigma_s) = \sigma_p^T\mathcal{I}\sigma_s$
$\pi(\sigma,\sigma_s)$ denotes the happiness of an individual given a mixed strategy chosen by society   
\item  \textbf{Payoff Matrix}
Payoff matrix for both the players is the interaction matrix $\mathcal{I}$.
\end{itemize}
\subsection*{Interaction matrix}
\begin{figure}[t]
  \small{
\begin{tabular}{|c|c|c|c|c|c|c|c|c|}
    \hline
        & a    & b    & c    & d    & e   & f   & g    & h\\\hline
    1   & 0.9  & -0.5 &	0.5  & 0.3  & 0.3 & 0.7 & 0.5  & -0.2\\\hline
    2  & 0.7  & 0.2  &	0	 & 0	& 0.4 & 0.7 & 0    & 0\\\hline
    3 & -0.1 & 0.8  &	-0.5 & -0.5 & 0   &	0	& -1   & 0\\\hline
    4  & -0.9 & 0.9  &	0    & 0    & 0.5 &	0.6 & 0    & 0\\\hline
    5   & 0.7  & 0.7  &	0    & 0.4  & 0.5 &	0.6 & 0    & 0\\\hline
    6  & -0.5 & 0	  & 0.8  & -0.9 & 0	  & 0   & 0.4  & 0.8\\\hline
    7 & 0.6  & 0.2  &	1	 & 0    & 0   &	0.5 & 0.8  & 0.5\\\hline
    8& 0    & 0.3  &	0    & 0.2  & 0	  & 0.5 & 0.3  & 0\\\hline
    9  & -0.5 & 0.5  &	0.5  & -0.8 & 0	  & 0.5 & 0.4  & 1\\\hline
    10   & 0    & 0.8  &	-0.2 & 0    & 0.2 &	0   & 0.3  & 0\\\hline
    11  & 0	   & -0.8 &	0.2	 & 0.5  & 0   &	0   & 0.4  & 0\\\hline
    12 & -0.4 & 0    &	-0.5 & 1	& 0   &	0   & -0.3 & 0.6\\\hline
    13& 0.2  & 0.2  & 0.5	 & -1   & 0   &	0	& -0.6 & -0.5\\\hline
\end{tabular}} \\
\centering
\end{figure}
This is an interaction matrix that we have used in our simulation where each cell represents the correlation between the column and the row property. It also represents the payoff matrix for both the population and the society as it weighs the positive and negative relationships between a strategy of and individual and society.\\

The above payoff matrix has 36 Nash equilibria in total. We have used an online calculator for Nash equilibria. This given the payoff matrix for both the agents calculates the Nash equilibria. Out of these 4 are pure strategy equilibria while the rest are mixed-strategy equilibria.


\section*{Protocol}
\subsection*{Simulation setting}
\begin{enumerate}
    \item We simulate each person of the population as having certain characteristics(strategies) and we impart a starting population to the city.
    \item Happiness $= \sum_{i=0}^N X_i^T\mathcal{I}\mathcal{C}$
    where $x_i$ is the characteristics of the $i^{th}$ person, $\mathcal{I}$ is the interaction matrix and $\mathcal{C}$ is the city characteristics.
 \end{enumerate}
 
\subsection*{Assumptions}
\begin{itemize}
    \item The \textit{happiness(fitness)} of a person is determined at his birth and remains constant during his lifetime.
    \item The \textit{lifespan} of a person is determined at his birth and is a function of his happiness.
    \item The \textit{mating frequency} of a person as well as the success of a \textit{mating} is a function of his happiness.
\end{itemize}

\subsection*{Algorithm}
\begin{algorithm}[H]
      \begin{algorithmic}[1]
        \STATE $\mathcal{P}\gets [X_1\cdots X_n] \sim \mathcal{N}(\mu, \Sigma)$
        \STATE $[(Y_1\cdots Y_k), (Z_1\cdots Z_k)] \gets Available(\{X_i\}_{i=1\cdots n})$
        \WHILE{$True$}
        \STATE $\{(Y_1, Z_{p_i})\}_{i=1\cdots k} \gets \underset{p\in \Pi}{argmax}\sum_{i=0}^k f(\theta_t, born(Y_1, Z_{p_i}))$
        \STATE $\{\hat{X_i}\}_{i=1\cdots k}\gets born(\{(Y_1, Z_{p_i})\})_{i=1\cdots k}$
        \STATE $\mathcal{P}\gets \text{Update\_pop}(X_1\cdots X_n, \hat{X_1}\cdots \hat{X_k})$
        \STATE $\theta_{t+1} \gets \theta_t + \Delta f(\theta_t, P)$
        \STATE $t \gets t + 1$
        \ENDWHILE
      \end{algorithmic}
      \caption{Pseudocode for simulation of city}
      \label{alg:seq}
\end{algorithm}

\subsection*{Timeline events}
\subsubsection*{Available}
Every person gets available for mating after a fixed time period called the \textit{Mating Gap}. After a few years, the males and females available among $\{X_i\}_{i=1\cdots n}$ form $[(Y_1\cdots Y_k), (Z_1\cdots Z_k)]$

\subsubsection*{Born}
\begin{figure}[H]
  \centering
  \includegraphics[width=3in]{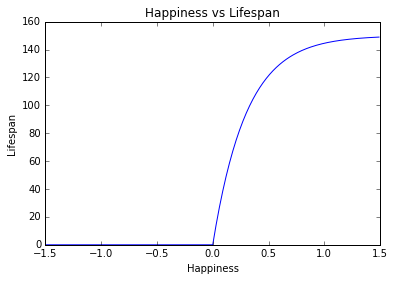}
  \caption{Lifespan vs happiness where $L = a * (1 - b e^{-h}))$ and $L$ is lifespan, $a=150$, $b=10$ and $h$ is defined as happiness}
  \label{lifespan}
\end{figure}

\begin{figure}[H]
  \centering
  \includegraphics[width=3in]{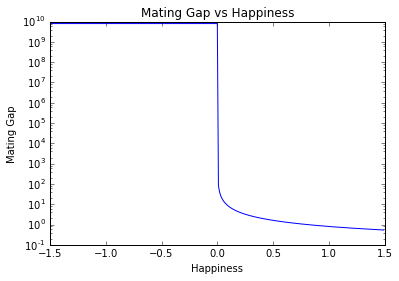}
  
  \caption{Mating Gap vs Happiness, $g = a / (clip(h ,0 , \infty ) + \epsilon)$ where $g$ is the mating gap, $a = 0.8$ and $h$ is defined as happiness.}
  \label{mating_gap}
\end{figure}

\begin{figure}[h]
  \includegraphics[width=3.5in]{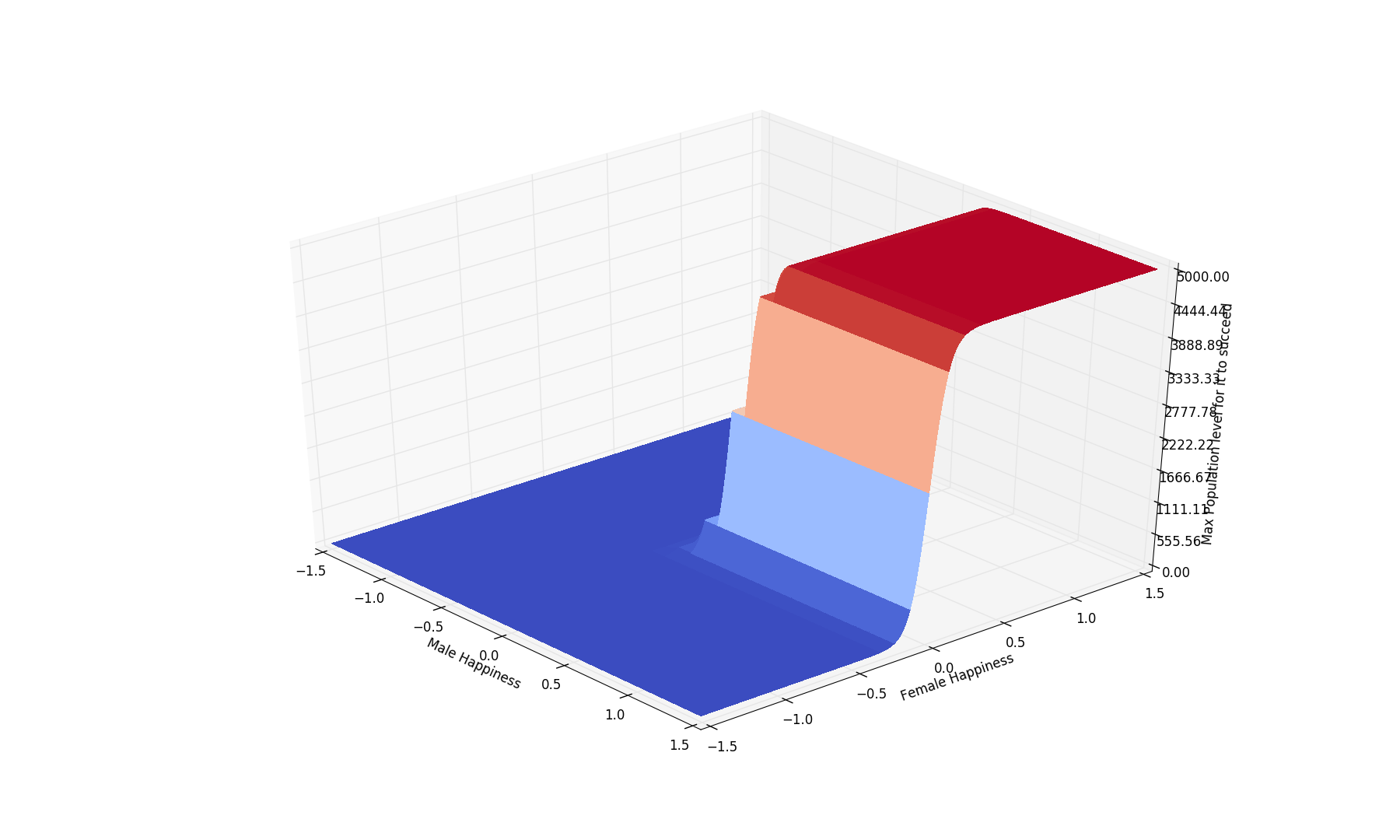}
  \caption{Mating success : This shows the min happiness of the male and the female required to succeed with respect to the current population of the city. $m =  a\cdot p +  max(1 - \sigma(scale * p_h), 1 - \sigma(scale * q_h))$ where $m$ is the min happiness, scale= 20 and $a=0.002$}
  \label{mate_success}
\end{figure}

\begin{itemize}
    \item \textbf{Uniform Selection} A child gets its characteristics from either its parents with uniform probability.
    \item \textbf{Mutation} A child mutates its genes to a uniform number with probability $p=0.1$
\end{itemize}
Figures[ \ref{lifespan}, \ref{mating_gap},\ref{mate_success}] show below our formulation of the three term along with how they vary with happiness of the individual.     

\section*{Optimization steps}
As the city's population evolves, both the population and city characteristics need to be updated accordingly. Intuitively, happier individuals should be able to survive for a longer time and also mate more frequently. These individuals would also then look for individuals to mate who can increase their happiness quotient. To solve this problem, we perform a bipartite matching with an added gaussian noise that accounts for the mismatches and makes the scenario more realistic. For city, the characteristics should update so that they support the population and hence increase their happiness. Therefore, we do a gradient ascent for updating the city characteristics. The detailed updates are mentioned below.

\subsection*{Population characteristics updation}
Given $\mathcal{Y}$ and $\mathcal{Z}$, which represent the available males and females and $\theta_t$, which represent the city characteristics, we need to find a permutation $p$ such that it solves the following optimization problem.
$$\underset{p\in \Pi}{argmax} \sum_{i=0}^k f(\theta_t, born(\mathcal{Y}_i, \mathcal{Z}_{p_i})) $$
where
$$f(\theta_t, \hat{X}) = \hat{X}^T(\mathcal{I}\theta_t)$$
This is a maximum weight bipartite matching problem also known as the \textit{assignment} problem.

\subsection*{Society characteristics updation}

Given a certain population characteristics $x = \frac{1}{N} \sum_{i=0}^N X_i$ and a current city characteristics, we need to  update the city characteristics such that the following two properties are satisfied:
\begin{itemize}
    \item The change is local i.e. a society doesn't undergo drastic changes in its characteristics overnight.
    \item The society must change in order to make its population happier.
\end{itemize}
Note that a simple way to perform this update is to perform gradient ascent.
$\theta_{t+1}\gets \theta_t + \lambda x^T\mathcal{I}$


\section*{Experiments and Results}

\subsection*{Population Survival}
In these experiments we have tried to look at how population of a city and it's net happiness evolves, depending upon the characteristic initialisation of the city and the characteristic initialisation of it's people. We have also obtained the plots for Population for the city over various years. We have defined the characteristic vectors to represent the various kinds of citirs and populations but we do not report them here for lack of space and will be included in an extended version of the paper.

\begin{figure}[H]
  \centering
  \subfloat[a]{ \includegraphics[width=.45\linewidth]{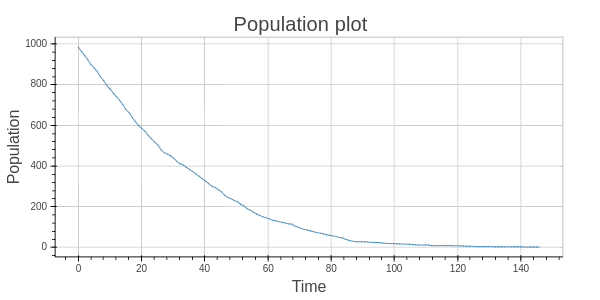}}\hfill
  \subfloat[b]{\includegraphics[width=0.45\linewidth]{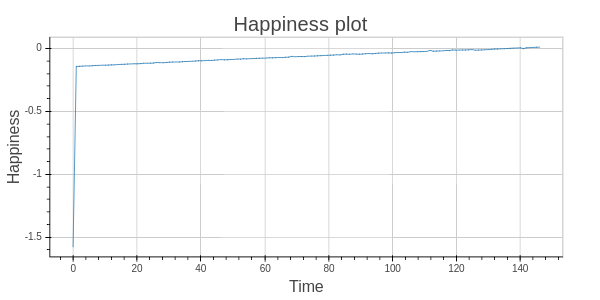} }\\
  \subfloat[c]{\includegraphics[width=0.45\linewidth]{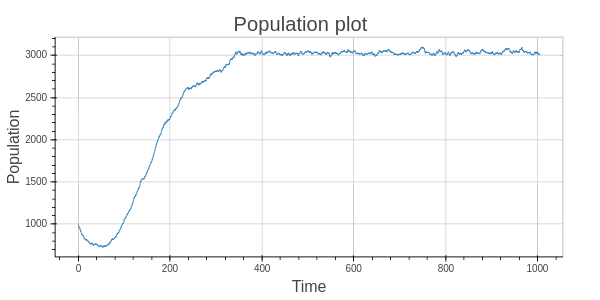}}\hfill
  \subfloat[d]{\includegraphics[width=0.45\linewidth]{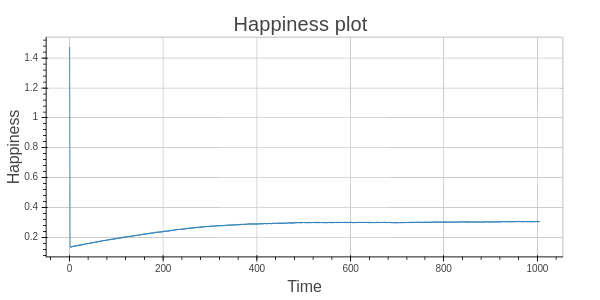} }
  \caption{Population and happines plots for a high intellect and a low intellect poulation in a criminal city. Figure(a) shows the population plot for an initial high intellect population in a criminal city. Figuer(b) shows the happines of the same population. Figure(c) and (d) shows the same thing for an initial criminal population in the same city}
  \label{fig:high-intel-low-intel-crim}
\end{figure}

\begin{figure}[H] 
    \centering
    \includegraphics[width=.48\linewidth]{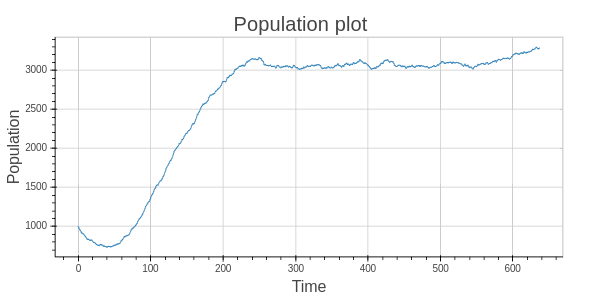} 
    \includegraphics[width=.48\linewidth]{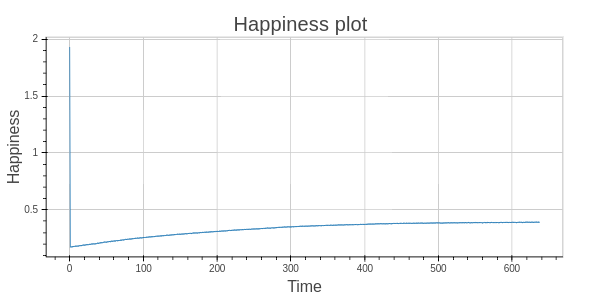} 
    \caption{High Intellect Population} 
    \includegraphics[width=.48\linewidth]{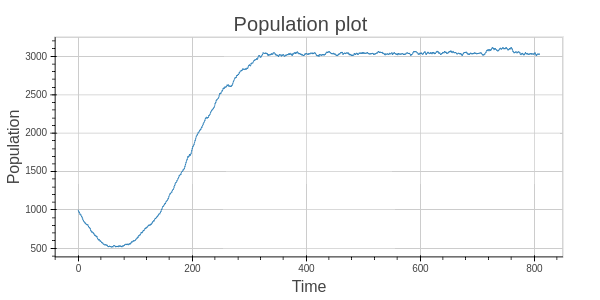} 
    \includegraphics[width=.48\linewidth]{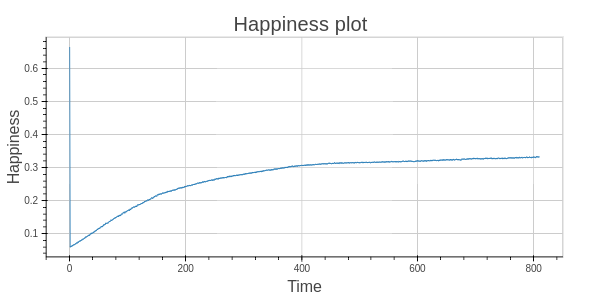} 
    \caption{Low Intellect Population}
    \label{fig:city_high_intel}
\end{figure}

The plots below(Figure[\ref{fig:high-intel-low-intel-crim}]) indicate how the various populations survived in a criminal city.
This is the same set of experiments for a intellectual city(Figure[\ref{fig:city_high_intel}]).

A very interesting thing to note in this plots is that the population seems to go on the verge of extinction and then recovers dramatically from it. We visualize this as a test for the population where only the fit people are able to survive. One can also notice that this is the point of time, from when the happiness starts to rise again. This refers to the initial difficult times a society has to suffer before it can actually flourish.

\subsection*{Non Optimal mating}
There are a series of more realistic experiments below which were done after feedback from our presentation. The most important change is to make the mating process more realisitic by removing the optimal matching algorithm we had used previously. Here we use the same matching algorithm i.e. maximal weight bipartite matching algorithm on the bipartite graph with the weight of edges modified by adding a gaussian noise with zero mean and unit variance. In order to simulate locality based maring, we take a more pessimistic view by randomly partitioning the people and then finding the optimal matching(with noise) in that partition. The new setting is as follows.

Given $\mathcal{Y_i}$ and $\mathcal{Z_i}$, which represent the available males and females of the $i^{th}$ partition and $\theta_t$, which represent the city characteristics, we need to find a permutation $p$ such that it solves the following optimization problem.
$$\underset{p\in \Pi}{argmax} \sum_{j=0}^k f(\theta_t, born(\mathcal{Y_i}_j, \mathcal{Z_i}_{p_j})) $$
where
$$f(\theta_t, \hat{X}) = \hat{X}^T(\mathcal{I}\theta_t) + z$$ and $$z ~\sim \mathcal{N}(0, 1)$$

Surprisingly, the results we obtained were better than the previous case in the following respects.
\begin{itemize}
\item The drop in population, which is observed initially is not that drastic as observed initially.
\item The convergent happiness is higher than the previous case.
\end{itemize}


\paragraph{Analysis}

\begin{figure}[t]
\centering    \subfloat[a]{\includegraphics[width=.48\linewidth]{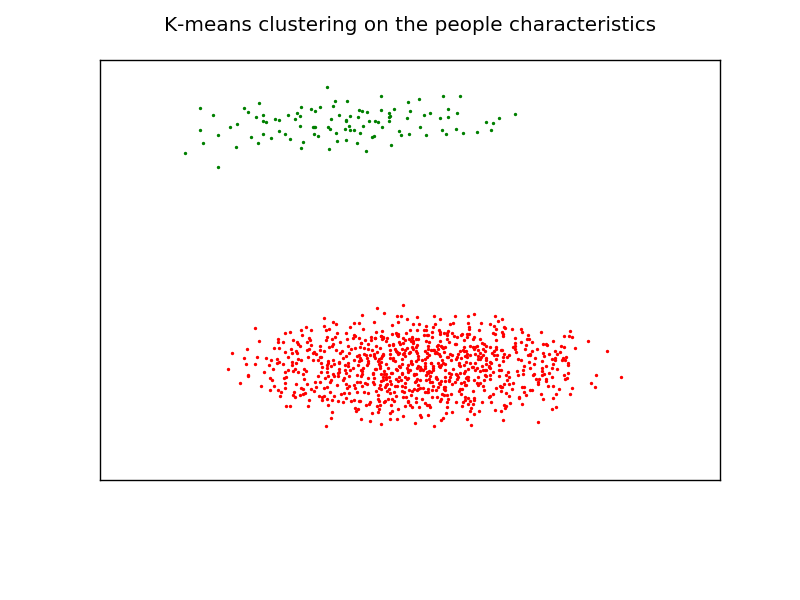}} \hfill \subfloat[b]{\includegraphics[width=.48\linewidth]{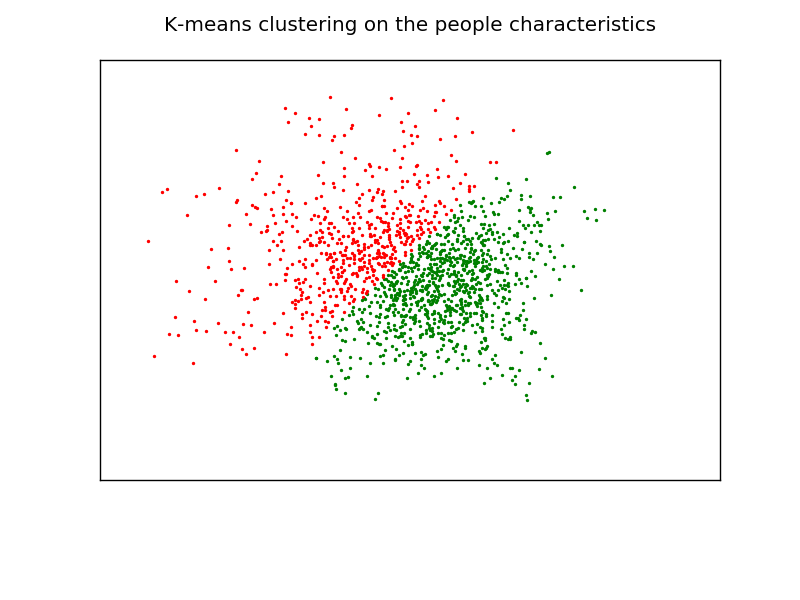} }\\ \subfloat[b]{\includegraphics[width=.48\linewidth]{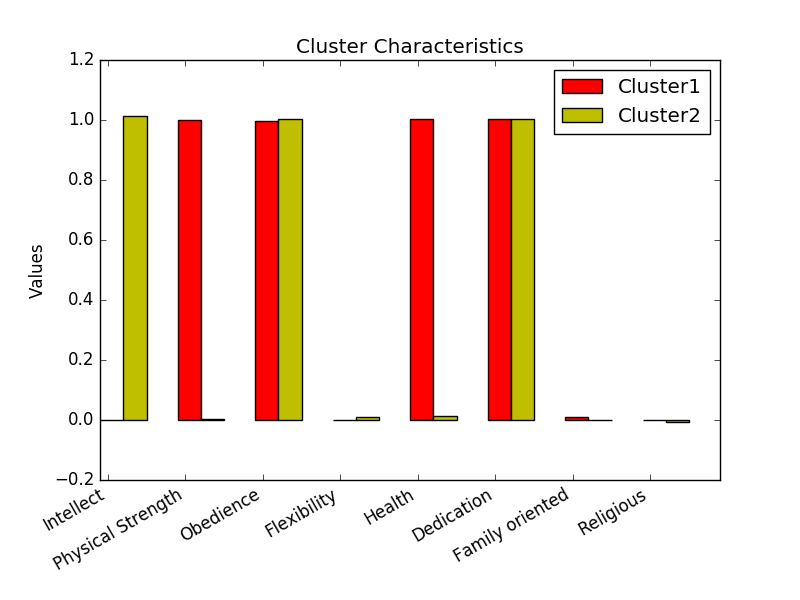} }\hfill \subfloat[d]{\includegraphics[width=.48\linewidth]{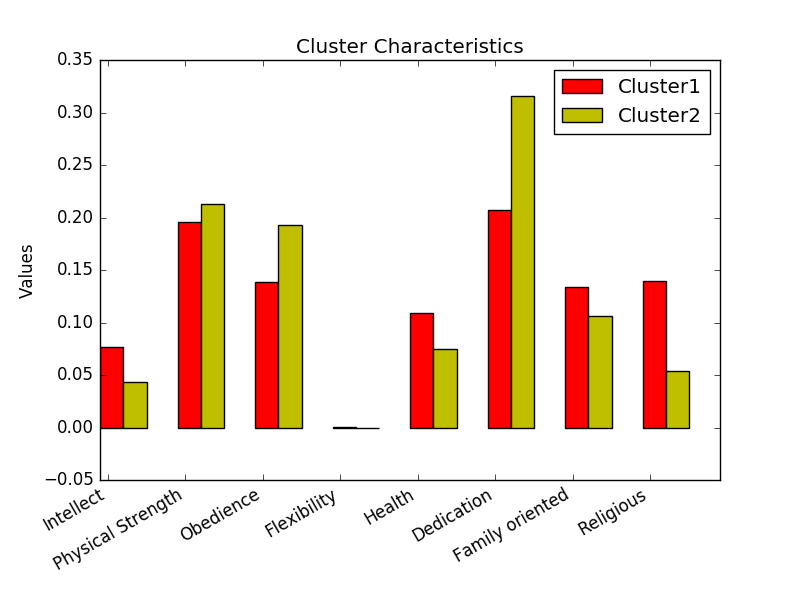}} 
   \caption{These experiments deal with a setting where there are $80\%$ Farmers and $20\%$ Intellectuals in Agrarian city
 (a): Initial Population Characteristics,(b): Final Population Cluster,(c): Final Population Characteristics,(d): Final Population Cluster} 
\label{sec:80-farmers-20}
\end{figure}

\begin{figure}[H]
  \subfloat[a]{\includegraphics[width=.45\linewidth]{./images/analysis_plots/initial_people_9farmers1highintellect_character.png}}\hfill
  \subfloat[b]{\includegraphics[width=.45\linewidth]{./images/analysis_plots/initial_people_9farmers1highintellect_cluster.png} }\\
 \subfloat[c]{\includegraphics[width=.45\linewidth]{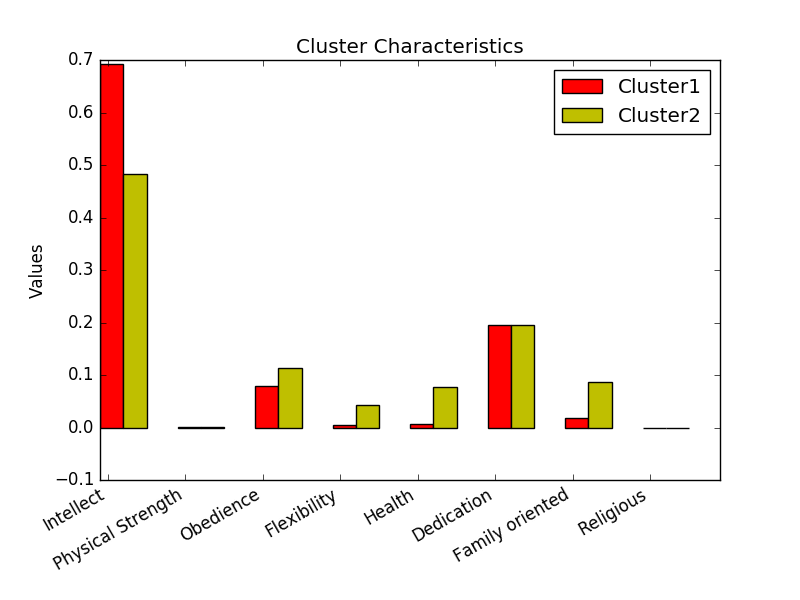} }\hfill
  \subfloat[d]{\includegraphics[width=.45\linewidth]{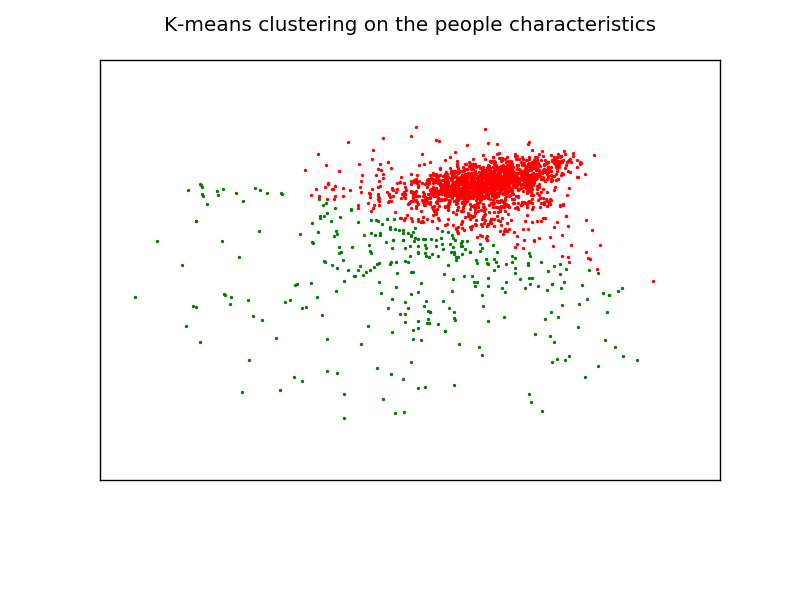} }
  \caption{Here there are $75\%$ Farmers $25\%$ High Intellect in a high intellect city. Figure(a) shows the Initial Population Characteristics, Figure (b) shows the Initial Population Clusters, Figure (c) shows the Final Population Characteristics and Figure (d) shows the Final Population Cluster}
  \label{sec:75-farmers-25high}
\end{figure}

\begin{figure}[H]
    \centering
    \subfloat[a]{ \includegraphics[width=.45\linewidth]{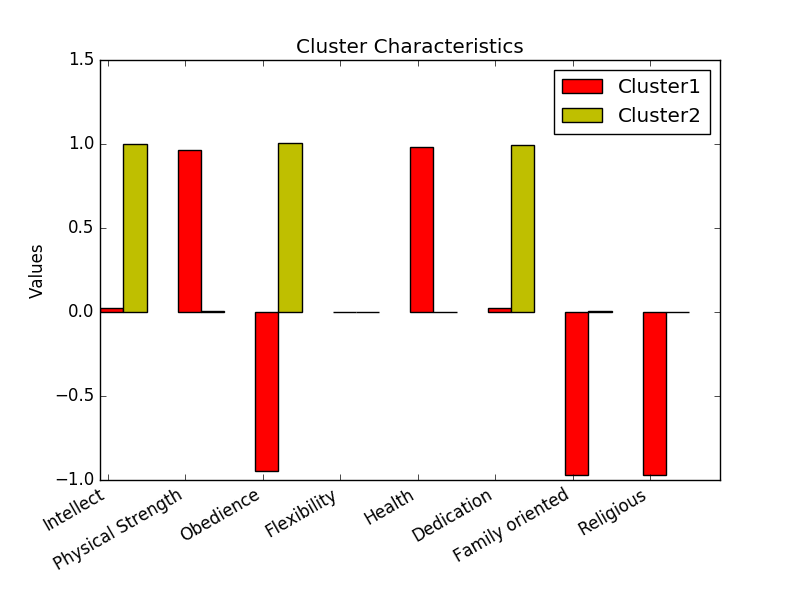} }\hfill \subfloat[b]{\includegraphics[width=.45\linewidth]{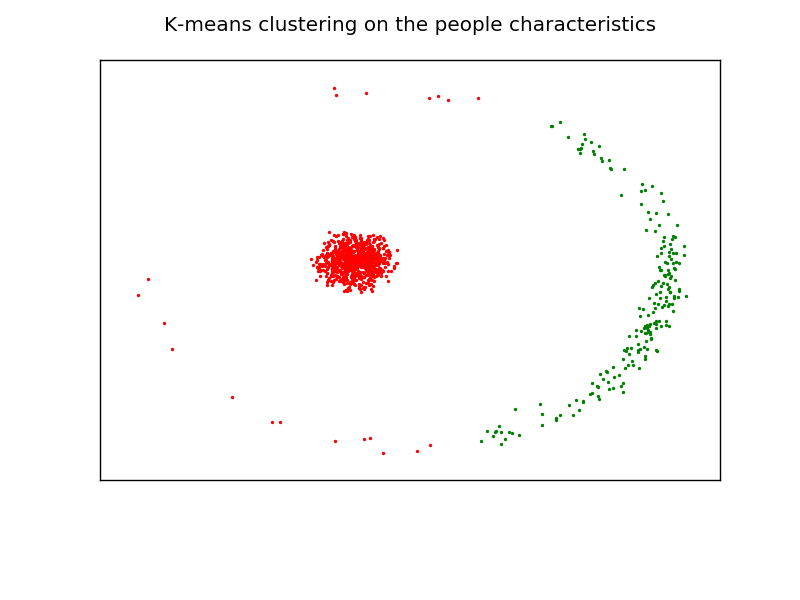}}\\
    \subfloat[c]{\includegraphics[width=.45\linewidth]{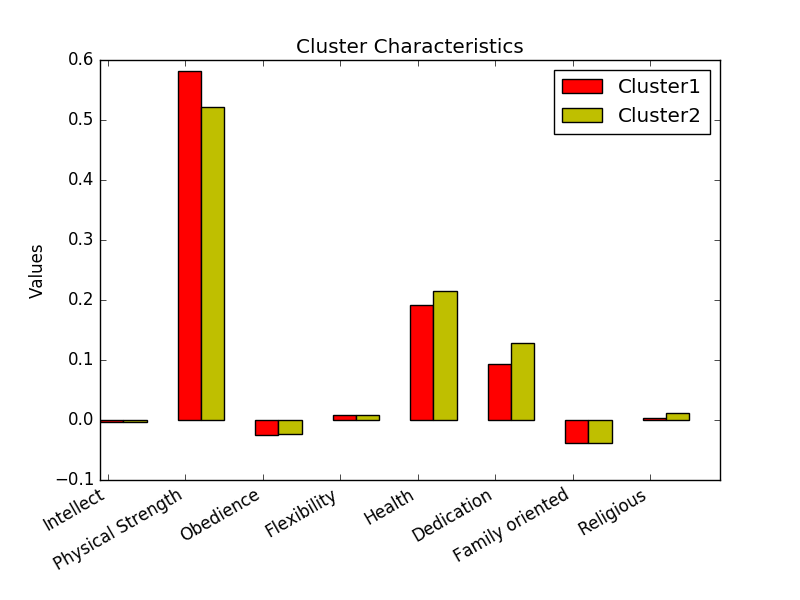} }\hfill
    \subfloat[d]{ \includegraphics[width=.45\linewidth]{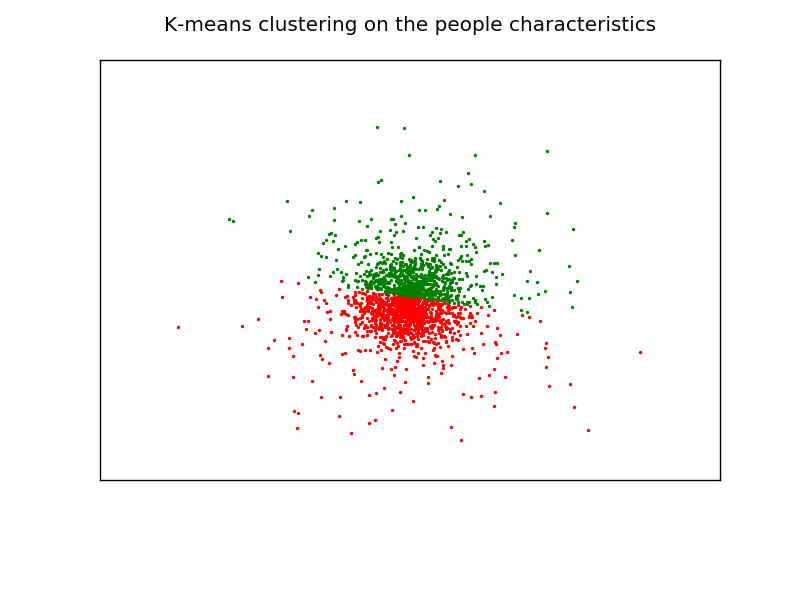} }
    \caption{Here there are $75\%$ Criminal $25\%$ High Intellect in a  criminal city. Figure(a) shows the Initial Population Characteristics, Figure (b) shows the Initial Population Clusters, Figure (c) shows the Final Population Characteristics and Figure (d) shows the Final Population Cluster}
    \label{sec:75-criminal-25-high-crim}
  \end{figure}

With this we conducted a new set of experiments, where we have a mixed initial population and after running the simulation for 10000 unit times, we again get the final population. We then apply \textbf{MDS or Multidimensional Scaling} to compress the data to two dimensions and then apply \textbf{K-Means} clustering to cluster the people to two clusters. After this, we analyze the characteristics of the initial cluster and the final clusters. We carry out three different experiments in three setting is Figure [\ref{sec:80-farmers-20}, \ref{sec:75-farmers-25high}, \ref{sec:75-criminal-25-high-crim}]. Below, we give our analysis in two of these settings.
In Figure[\ref{sec:80-farmers-20}], it is noticeable how the population is no more highly distinguishable after the simulation runs. This is evident from the K-Means clustering shown. One can also notice how the agrarian city forces the people to have uniformly high physically strength and not so high intellect. Health, which was another point of difference between the two populations have also grown to a somewhat more uniform distribution after the simulation.

  In Figure[\ref{sec:75-farmers-25high}], again the most noticeable feature is how the population grows to be unimodal in its characteristics after the simulation is run. Being an intellectual city, the most important characteristics i.e. intelligence grows after the simulation to a remarkable extent while the other characteristics remains more or less uniform. The other important thing is dedication which remains at a somewhat higher level than the other characteristics.

  In Figure[\ref{sec:75-criminal-25-high-crim}], the experiment suggests that the final population need not follow one of its founding sub categories but may indeed develop new characteristics to survive. Here, health and obedience were initially present in different clusters at a high level. However, in the final phase, it is present equally in both the clusters. On the other hand , religiousness, which is a quality favoured by both the dominant seed population as well as the city has seemed to grow the other way i.e. positive. Family oriented and obedience though present in high quantity initially and favoured by the  city has grown to stabilize at a more moderate level finally.

  \subsection*{Locality based mating}
  In an attempt to make our simulation more realisitic, we also include the factor of locality into the mating selection procedure. This means that while mating, one not only looks at the expected happiness of their child but also at the possibility of the matching itself depending upon the how far the two people are situated. This means that added to the already existing personal traits, we now have $(x,y)$ coordinate of each person/ city block to which the person belongs. After the birth of the child, the new child belongs to either one of the city blocks(father's or mothers) randomly. Hence, the objective function of the mating process is as follows
  $$\underset{p\in \Pi}{argmax} \sum_{i=0}^k f(\theta_t, born(\mathcal{Y}_i, \mathcal{Z}_{p_i}))  - \gamma d(\mathcal{Y}_i, \mathcal{Z}_{p_i})$$ 
where
$$f(\theta_t, \hat{X}) = \hat{X}^T(\mathcal{I}\theta_t)$$
and $\gamma$ is a scaling factor and $d$ is the distance function. For simplicity, we use the hamming distance.
\begin{figure*}[h]
  \centering
  \subfloat[a]{\includegraphics[width=.3\linewidth]{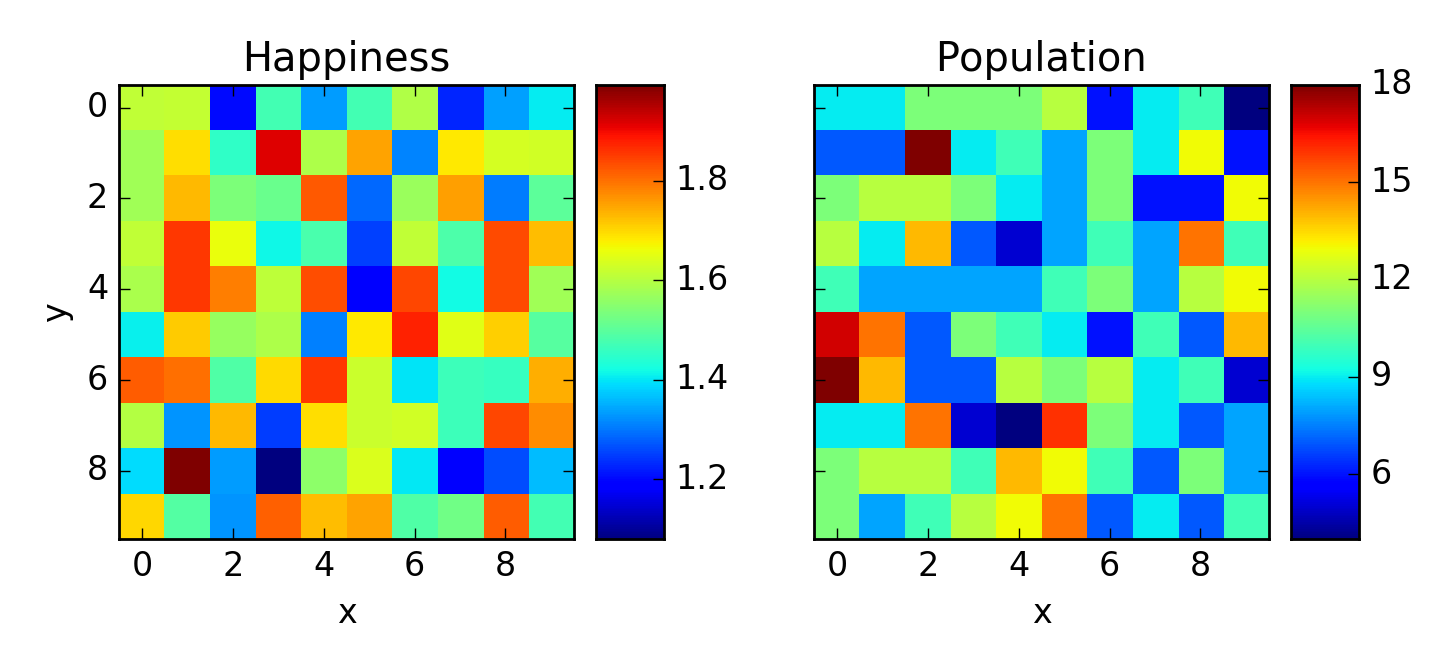}}
  \subfloat[b]{\includegraphics[width=.3\linewidth]{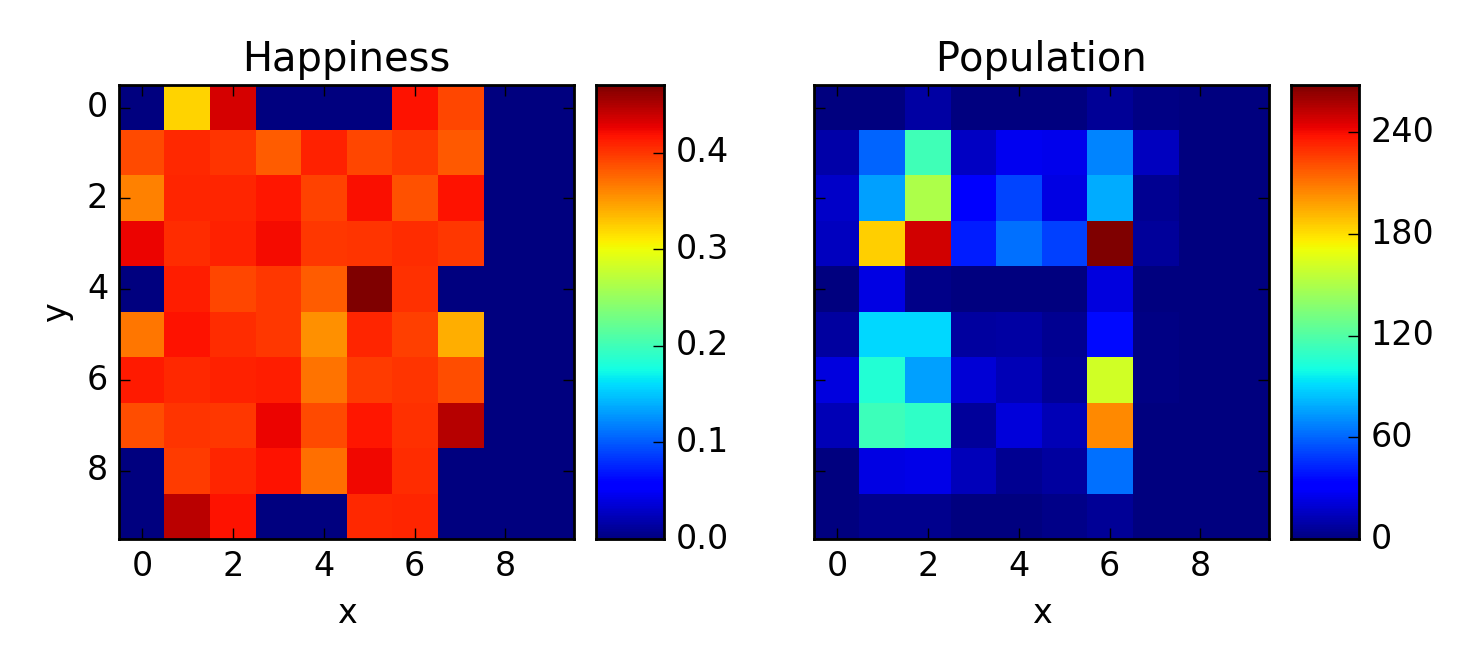}}
  \subfloat[c]{\includegraphics[width=.3\linewidth]{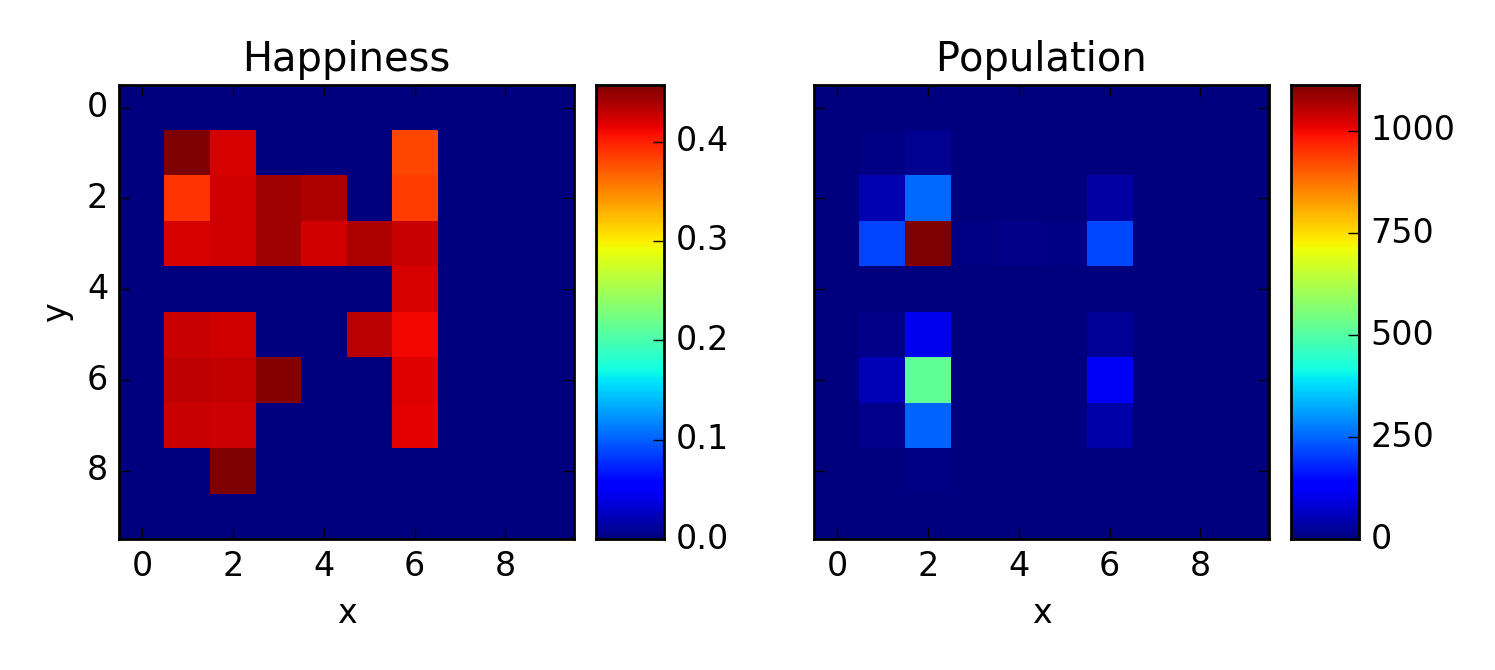}}
  \caption{Locality based mating. Figure(a) represents the initial distribution and Figure(b) represents the final distribution. In each plot, the figure on the left hand size represents the average happiness and the figure on the right hand side represents the population of that grid block. Here the grid contains 100 blocks(10x10)}
  \label{fig:locality}
\end{figure*}

In Figure[\ref{fig:locality}], we plot the distribution of the population and average happiness of the population initially and finally(i.e. after the simulation). We find that multiple communities evolve in the space and some of these communities are disconnected from each other completely. However, all of these communities maintain a viable population. Another possible extension to this is reducing the dependence on the total population(in mating success) to the total population in a particular grid as opposed to the total population of the entire society.

\subsection*{Learning Rate}
\subsubsection*{Variations of Learning Rate}
\begin{figure}[H]
  \centering
  \subfloat[1X]{\includegraphics[width=.45\linewidth]{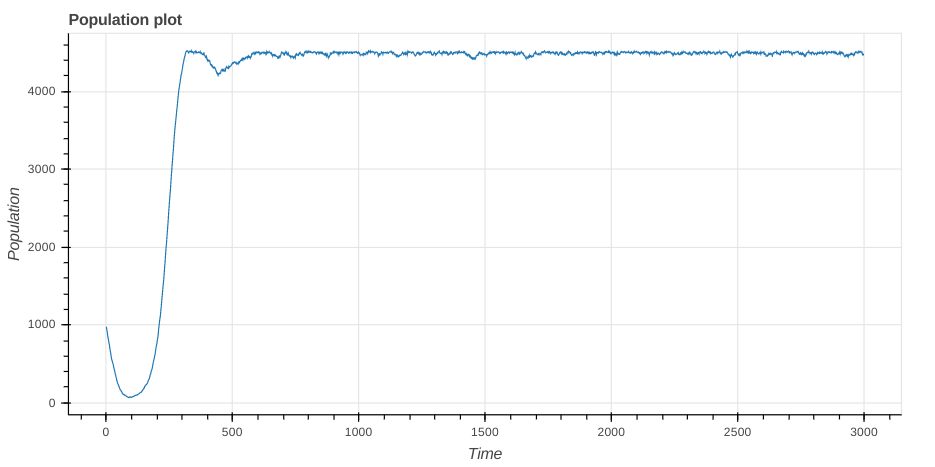}}\hfill
  \subfloat[3X]{\includegraphics[width=.45\linewidth]{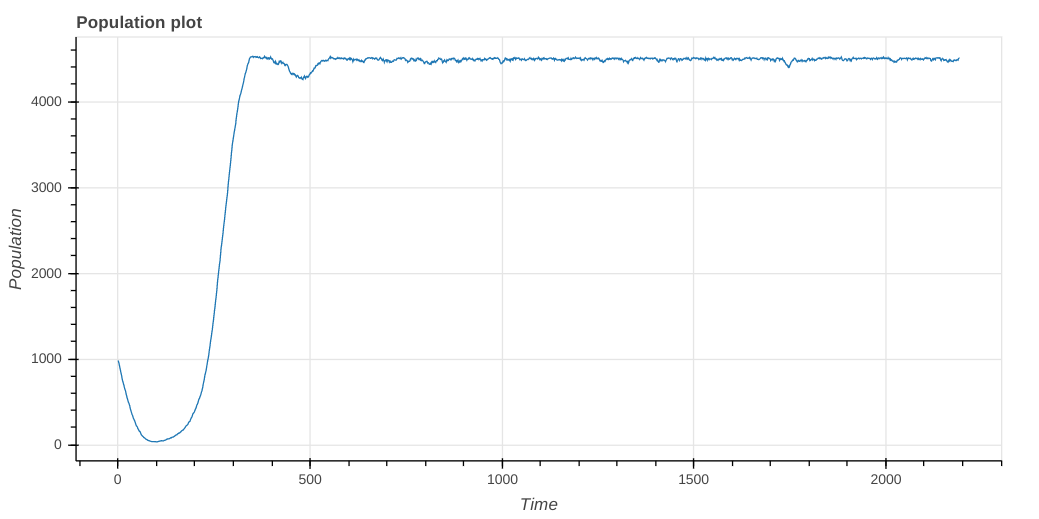}}\\
  \subfloat[10X]{\includegraphics[width=.45\linewidth]{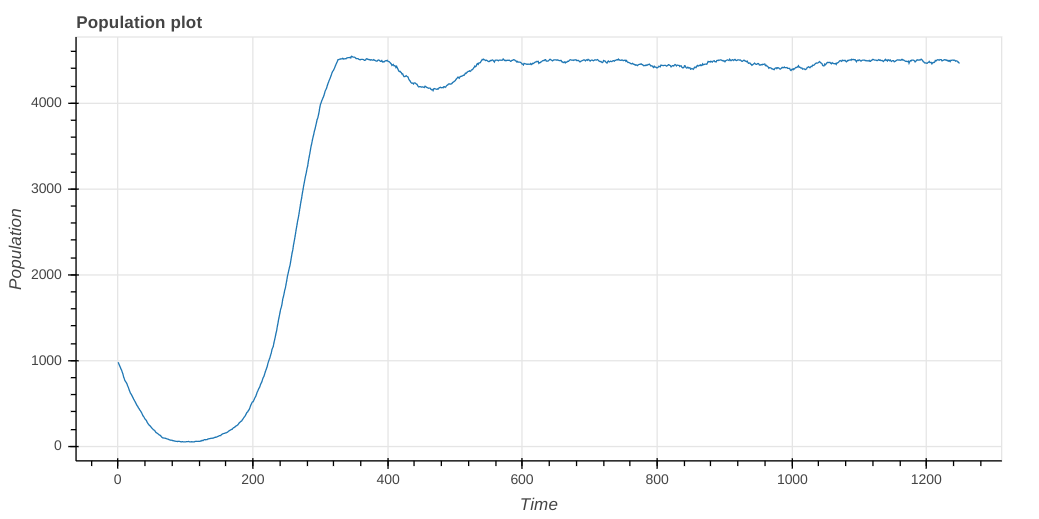}}\hfill
  \subfloat[30X]{\includegraphics[width=.45\linewidth]{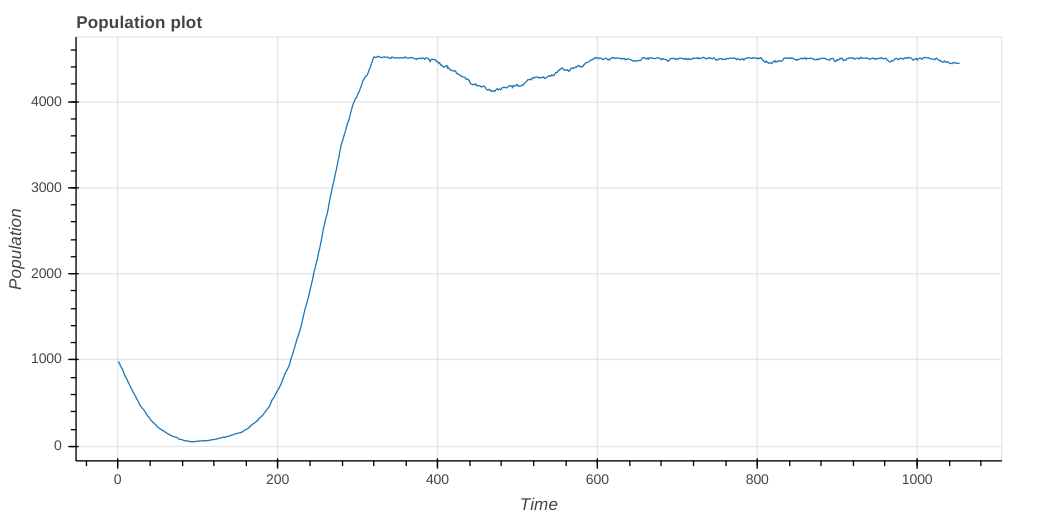}}
  \caption{Population}
  \label{fig:lambdaPop}
\end{figure}

\begin{figure}[H]
  \centering
  \subfloat[1X]{\includegraphics[width=.45\linewidth]{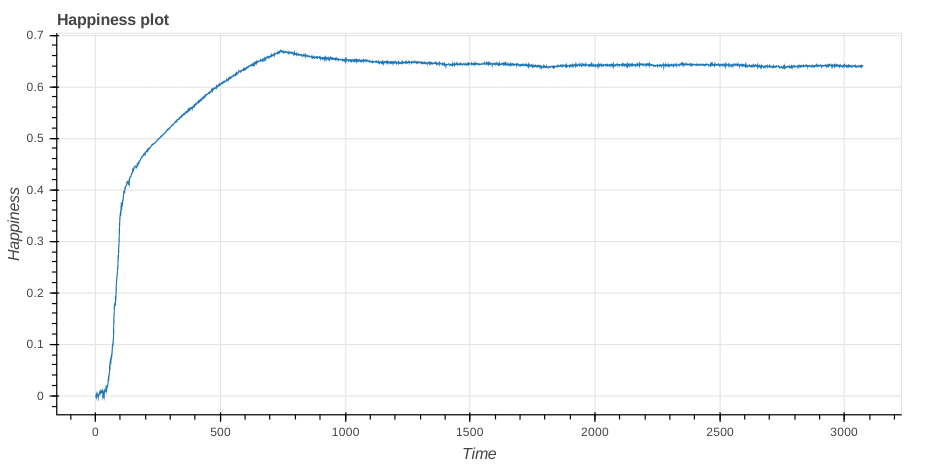}}\hfill
  \subfloat[3X]{\includegraphics[width=.45\linewidth]{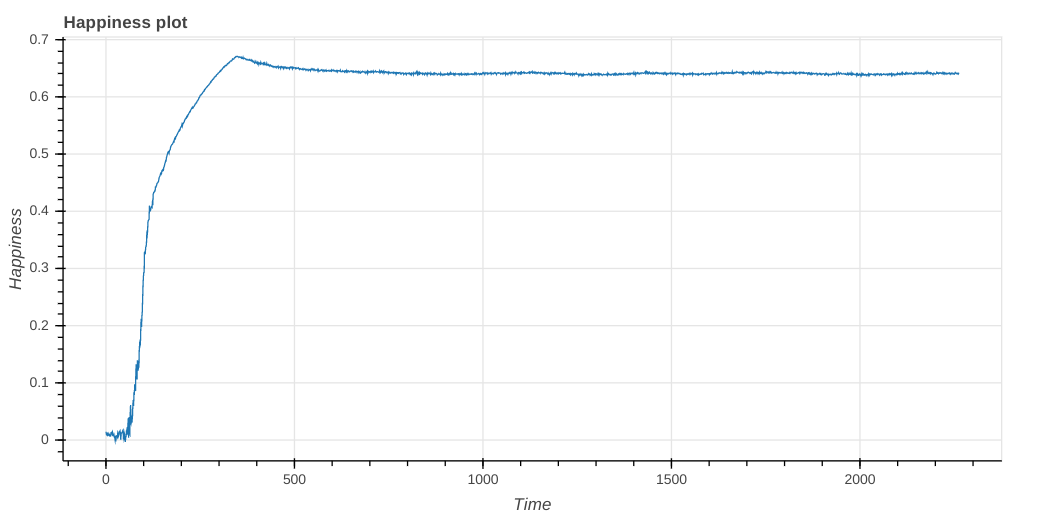}}\\
  \subfloat[10X]{\includegraphics[width=.45\linewidth]{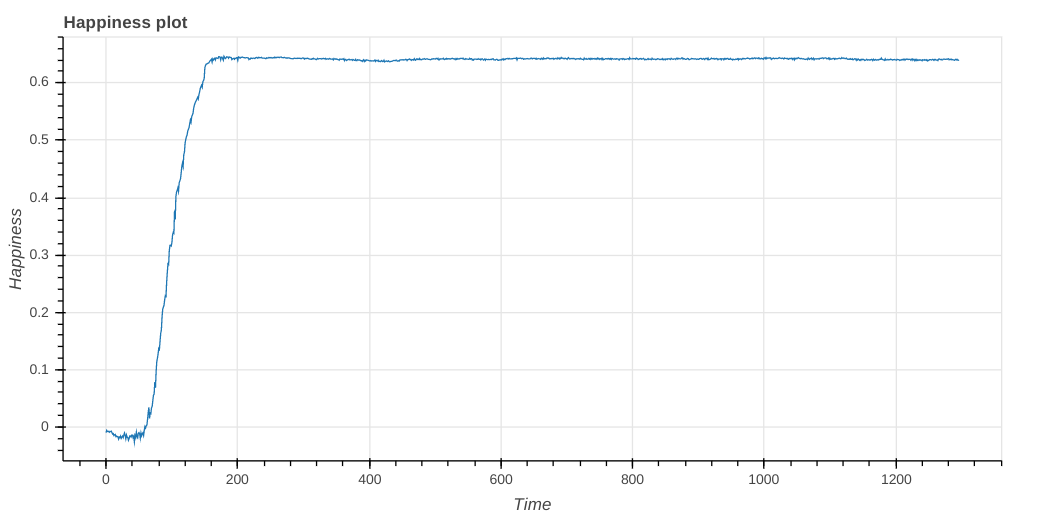}}\hfill
  \subfloat[30X]{\includegraphics[width=.45\linewidth]{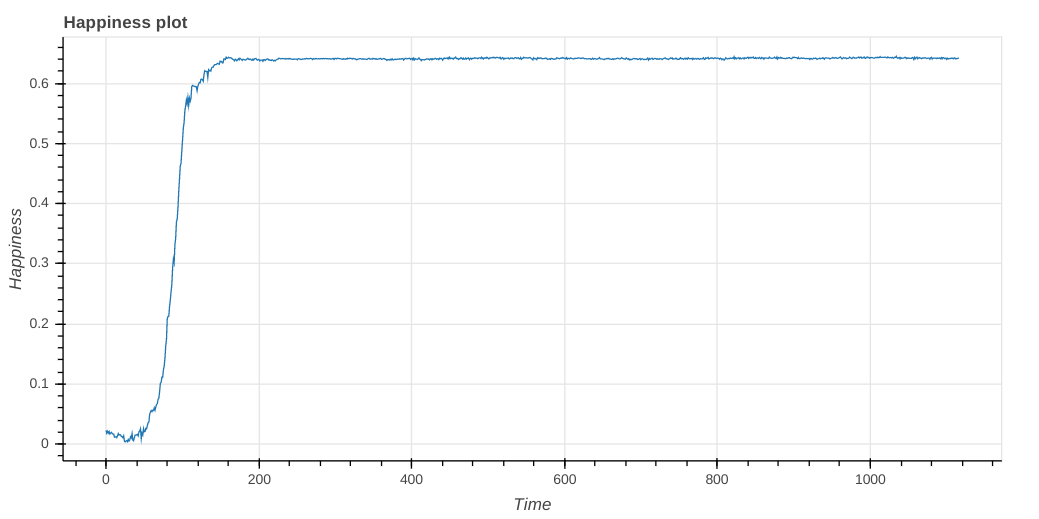}}
  \caption{Happiness}
  \label{fig:happ}
\end{figure}

Learning rate controls the rate of change of societal characteristics which happens through the gradient ascent Step. We denote Learning rate by 
$\lambda$ .
We can adjust $\lambda$ and see the changes it causes.We initialise $\lambda$ at $10^{-4}$ and then vary it in multiples of 1,3,10 and 30 of the original value.
Here we report the evolution plots(Figure:[\ref{fig:lambdaPop}, \ref{fig:happ}]) of population which are obtained by varying $\lambda$ respectively across multiples of 1,3,10 and 30. We report the population
and happiness plots seperately.

In the above plots(Figure:[\ref{fig:lambdaPop}, \ref{fig:happ}) we see that varying lambda affects the time which takes in evolution of happiness levels in society. Higher lambda changes societal
characteristics fast and achieves the stable happiness values earlier. The happiness value eventually plateau.

\subsubsection{Feature Sets for these experiments}
For these sets of experiments we experimented with a varied set of features than mentioned above, for both individuals as well as the Society.
For individuals we took 3 features from the Big Five Personality traits model. Namely Openness to Experience, Conscientiousness and Extraversion.
We combined them with IQ and Physical Strength to describe an agent in our model. For societal characteristics we have 5 features, namely, Intellectuality in the society, Living Standard,
Crime Rate, Industrialisation and Cultural Richness of society. We heuristically filled an interaction matrix using corelation values between the Individual and Societal Characteristics to complete our model.

\subsection{Dynamic Learning Rate}
We also propose Dynamic Learning Rate to more realistically model the evolution procedure. Rate of evolution depends on the flexibility of people towards change.
Thus there are various Personality Traits, for example Openness to Experience, which encodes how flexible a person is towards new values and experiences.
Thus we propose to use average of this trait(or combination of such traits) over the whole population to dynamically determine the value of $\lambda$ at any point of time.
This understanding of learning rate is meant to model the dependence of rate of evolution which depends upon the liberality of people.



\section*{Contribution and future work}
We have built a framework to carry out simulations on this protocol in a very efficient way using the \textit{simpy} library, which offers the capacity of simulation of multiple events in different timescales in an event based callback type of framework. We also offer real time logging capacity of all the characteristics of the population as well as the population strength and the happiness. Our main contribution is that we  propose an algorithm to explain the development of social characteristics of a population, which uses two different kind of optimization algorithms, which are well suited to the particular cause as needed here. It must also be noted that both of these steps are cheap and hence easy to simulate. By viewing it as a two player evolutionary game, further work could include drawing similarity to the fictitious play protocol.

We also offer an alternate explanation to the existence of this second player called \textit{society} in our game. The happiness of a society can be described as the average happiness of all $ \binom{n}{2}$ pairs of people. This requires $O(n^2)$ computation. However, by assuming the existence of a latent variable called \textit{Society} to describe the net effect of the other people in the interaction , this can be done in $O(n)$ time. Alternate maximization technique can here lead to optimizing the net happiness as well as bring the city more close to this proposed latent variable.

However, this does not necessarily mean that the city characteristic serves no other purpose than to model the $O(n^2)$ interaction in an $O(n)$ representation. It also seves to model the environmental and social constraints faced by the population e.g. existing farm lands in huge quantity, an economy dependant on a certain trade, abscence of fertile lands, rule of a certain kind of government/prevalence of certain norms or laws. This does not necessarily mean that they cannot change. It only means that the rate of change is relatively less and it might actually force the model to converge in a certain local minima, which might respect these constraints.

\bibliographystyle{IEEEtran}
\bibliography{masbib}

\begin{thebibliography}{1}
\providecommand{\url}[1]{#1}
\csname url@samestyle\endcsname
\providecommand{\newblock}{\relax}
\providecommand{\bibinfo}[2]{#2}
\providecommand{\BIBentrySTDinterwordspacing}{\spaceskip=0pt\relax}
\providecommand{\BIBentryALTinterwordstretchfactor}{4}
\providecommand{\BIBentryALTinterwordspacing}{\spaceskip=\fontdimen2\font plus
\BIBentryALTinterwordstretchfactor\fontdimen3\font minus
  \fontdimen4\font\relax}
\providecommand{\BIBforeignlanguage}[2]{{%
\expandafter\ifx\csname l@#1\endcsname\relax
\typeout{** WARNING: IEEEtran.bst: No hyphenation pattern has been}%
\typeout{** loaded for the language `#1'. Using the pattern for}%
\typeout{** the default language instead.}%
\else
\language=\csname l@#1\endcsname
\fi
#2}}
\providecommand{\BIBdecl}{\relax}
\BIBdecl

\bibitem{darwin1951origin}
C.~Darwin and G.~Beer, \emph{The origin of species}.\hskip 1em plus 0.5em minus
  0.4em\relax Dent, 1951.

\bibitem{miller2009complex}
J.~H. Miller and S.~E. Page, \emph{Complex adaptive systems: An introduction to
  computational models of social life}.\hskip 1em plus 0.5em minus 0.4em\relax
  Princeton university press, 2009.

\bibitem{scott2000rational}
J.~Scott, ``Rational choice theory,'' \emph{Understanding contemporary society:
  Theories of the present}, vol. 129, 2000.

\bibitem{downs1994conflict}
G.~W. Downs and D.~M. Rocke, ``Conflict, agency, and gambling for resurrection:
  The principal-agent problem goes to war,'' \emph{American Journal of
  Political Science}, pp. 362--380, 1994.

\bibitem{simon1957models}
H.~A. Simon, ``Models of man; social and rational.'' 1957.

\end{thebibliography}
\end{document}